\documentclass[pra,twocolumn, notitlepage]{revtex4-2}
\usepackage{amsmath, amssymb, graphicx, bm}
\DeclareMathOperator{\sech}{\mathrm{sech}}
\DeclareMathOperator{\diag}{\mathrm{diag}}
\usepackage{qcircuit}

\usepackage[frozencache =true,cachedir=.]{minted}
\usepackage{xcolor}

\definecolor{LightGray}{gray}{0.9}

\begin{document}

\title{Fast differentiable evolution of quantum states under Gaussian transformations}

\author{Yuan Yao}
\affiliation{Institut Polytechnique de Paris}
\affiliation{T\'el\'ecom Paris, LTCI, 19 Place Marguerite Perey 91120 Palaiseau, France}

\author{Filippo M. Miatto}
\affiliation{Xanadu Quantum Technologies, 777 Bay St.~Toronto, Canada}
\affiliation{Institut Polytechnique de Paris}
\affiliation{T\'el\'ecom Paris, LTCI, 19 Place Marguerite Perey 91120 Palaiseau, France}

\date{\today}
\begin{abstract}
In a recent work we presented a recursive algorithm to compute the matrix elements of a generic Gaussian transformation in the photon-number basis. Its purpose was to evolve a quantum state by building the transformation matrix and subsequently computing the matrix-vector product. Here we present a faster algorithm that computes the final state without having to generate the full transformation matrix first. With this algorithm we bring the time complexity of computing the Gaussian evolution of an $N$-dimensional $M$-mode state  from $O(MN^{2M})$ to $O(M(N^2/2)^M)$, which is an exponential improvement in the number of modes. In the special case of high squeezing, the evolved state can be approximated with complexity $O(MN^{M})$. Our new algorithm is differentiable, which means we can use it in conjunction with gradient-based optimizers for circuit optimization tasks. We benchmark our algorithm by optimizing circuits to produce single photons, Gottesman-Kitaev-Preskill states and NOON states, showing that it is up to one order of magnitude faster than the state of the art. 
\end{abstract}

\maketitle

\section{Introduction}

Parametrized optical quantum circuits process quantum information by encoding it in quantum states of light and by propagating them through a sequence of parametrized optical gates \cite{peruzzo2014variational,flamini2018photonic,killoran2019continuous}. Our main goal is to design a fast, automated procedure to design  quantum optical circuits. Our focus is on the speed of the optimization: by using faster algorithms we can optimize larger devices, or similarly-sized devices to a higher precision.

Circuits in the optical domain are interesting because they enjoy of a few advantages over other technologies, such as no strict requirements for low temperatures or vacuum, and the potential of building compact and robust devices with integrated optics \cite{dietrich2016gaas, lenzini2018integrated,silverstone2016silicon}. Furthermore, optical quantum devices may be easily plugged into existing fiber networks for quantum communication purposes \cite{cozzolino2019high,muralidharan2016optimal}.

When using our algorithm, we model a quantum state propagating through an optical quantum circuit and we craft a differentiable loss function that evaluates how much the output of the circuit deviates from the target output. We then compute the gradient of the loss function with respect the parameters of the circuit and apply gradient descent to train the parameters until the output matches the target. After convergence we can read out the value of the parameters, effectively obtaining a design for a circuit that behaves as desired.

The circuit architecture that we consider is layered and reminiscent of dense neural networks, in that we alternate between a Gaussian (linear) transformation and a Kerr (non-linear) transformation. All Gaussian transformations are linear in the sense that they are always equivalent to a linear (affine) transformation of the optical phase space coordinates \cite{killoran2019continuous}. Note that there are ways in which Gaussian transformations can be considered non-linear, for example by the fact that their Hamiltonian can be non-linear in the quadrature operators. Our choice of architecture combines simplicity (each Gaussian+Kerr layer is functionally the same), and generality (we can implement any quantum transformation by stacking enough layers).

Parameterized optical circuits can already be trained via machine learning techniques for state preparation, gate synthesis, function simulation, image generation, hybrid classical-quantum classifiers, Hamiltonian simulation and quantum repeaters \cite{killoran2019continuous,arrazola2019machine,killoran2019strawberry,steinbrecher2019quantum}.
However, current methods can quickly become very demanding in terms of computational resources, and this makes it challenging to apply them to the design of devices that span several optical modes. Moreover, as the size of the Hilbert space grows, there can be vanishing-gradient effects that can impact the optimization of a parametrized circuit \cite{cerezo2020cost}.

Our solution in \cite{miatto2020fast} was up to 100 times faster than the previous state of the art, and in this work we increase the speedup exponentially in the number of modes $M$ by a factor $2^M$.
Here is a summary of the contributions presented in this work:
\begin{enumerate}
    \item We present an algorithm to directly compute the output state of a quantum optical circuit by computing only a fraction $1/2^M$ of tensor elements with respect to the whole transformation tensor. This exponential improvement benefits both the forward and the backward pass.
    \item Our algorithm is differentiable, as it relies on a linear recurrence relation.
    \item  Backpropagated gradient tensors have the shape of the state, rather than the shape of the transformation. Such quadratic improvement benefits the backward pass.
\end{enumerate}

This paper is organized as follows. In section 2 we recall the recurrence relation that was at the core of our previous algorithm, and that we will adapt for our new algorithm, which we present in section 3. In section 4 we first illustrate the backpropagation mechanism with complex parameters and explain how we obtain the gradient of a transformed state with respect to the parameters of the transformation. In section 5 we compare the runtime of the forward pass of our algorithm with respect to the current state of the art, and we benchmark the backward pass on a circuit optimization task. Finally, we discuss the results and future research directions in section 6.

\section{Preliminaries}

\subsection{Definition of an $M$-mode general Gaussian transformation}
Here we present the circuits that implement general Gaussian transformations, which are based on the Bloch-Messiah decomposition \cite{cariolaro2016bloch, cariolaro2016reexamination}.

In the single-mode case we use a sequence of squeezer, phase rotation and displacement gates:
\begin{align}
    \mathcal{G}^{(1)}(\gamma,\phi,\zeta) = D(\gamma)R(\phi)S(\zeta).
\end{align}
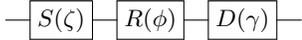
\begin{figure}[h]
$$
\Qcircuit @C=1em @R=.7em {
& \gate{S(\zeta)} & \gate{R(\phi)} &\gate{D(\gamma)} & \qw
}
$$
\caption{Single-mode Gaussian transformation architecture (states move through the circuit left to right).}
\label{fig:singlemodearch}
\end{figure}

In the two-mode case we have:
\begin{align}
    \mathcal{G}^{(2)}(\bm \gamma,\bm \phi,\theta',\varphi',\bm \zeta,\theta,\varphi) = D(\bm \gamma)R(\bm \phi)B(\theta',\varphi')S(\bm \zeta)B(\theta,\varphi).
\end{align}

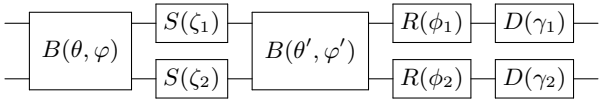
\begin{figure}[h]
$$
\Qcircuit @C=1em @R=.7em {
&\multigate{1}{B(\theta,\varphi)} &\gate{S(\zeta_1)}& \multigate{1}{B(\theta',\varphi')} & \gate{R(\phi_1)} & \gate{D(\gamma_1)}&\qw\\
&\ghost{B(\theta,\varphi)} & \gate{S(\zeta_2)} &\ghost{B(\theta',\varphi')} & \gate{R(\phi_2)} & \gate{D(\gamma_2)} &\qw
}
$$
\caption{Two-mode Gaussian transformation architecture.}
\label{fig:twomodearch}
\end{figure}

For $M$ modes, we have an $M$-mode interferometer, $M$ single-mode squeezers, a second $M$-mode interferometer and $M$ single-mode displacements. In this case, the interferometers need to be decomposed as a suitable arrangement of beamsplitters, for example following Clements decomposition \cite{clements2016optimal, reck1994experimental,de2018simple}.

\subsection{Notation and definitions}
We adopt a multi-index notation, i.e., $\mathcal{G}_{\bm k} = \mathcal{G}_{k_1,k_2,\dots}$, but we will show all indices when necessary. Transformations on $M$ modes are tensors of rank $2M$, where the first half of their indices correspond to the output and the second half correspond to the input (but we maintain the terminology ``matrix elements''). We also indicate as $\bm 0$ the vector of all zeros and by $1_i$ a vector with all zeros and a single 1 in position $i$.

Eq.~(29) from our previous work \cite{miatto2020fast} presents a recurrence relation to express the matrix elements of a general Gaussian transformation up to a global phase:
\begin{align}
& \mathcal{G}_{\bm 0} = C, \nonumber\\
\label{recrel}
& \mathcal{G}_{\bm k + 1_i} = \frac{1}{\sqrt{k_i+1}}
    \left(
    \mathcal{G}_{\bm k}\mu_i - \sum_{l = 1}^{2M}\sqrt{k_l}\mathcal{G}_{\bm k - 1_l}\Sigma_{il}
    \right).
\end{align}
Here $\mathcal{G}_{\bm k}$ denotes the matrix elements of a Gaussian transformation and $C,\mu$ and $\Sigma$ are defined as follows:

\begin{align}
&C = \frac{ \exp\left(-\tfrac{1}{2} \left[  ||\bm{\gamma}||^2  + \bm{\gamma}^\dagger  \bm{W} \diag(e^{i \bm{\delta}}  \tanh \bm{r}) \bm{W}^T  \bm{\gamma}^*\right]\right)}{\sqrt{\prod_{i=1}^M \cosh r_i}},\label{Eq:C}\\
&\bm{\mu}^T = \nonumber \\
& \left[\bm{\gamma}^\dagger \bm{W} \diag(e^{i \bm{\delta}}  \tanh \bm{r}) \bm{W}^T+\bm{\gamma}^T, -\bm{\gamma}^\dagger \bm{W} \diag(\sech  \bm{r}) \bm{V} \right],\label{Eq:mu}\\
&\bm{\Sigma} = \nonumber\\
&\left[	\begin{array}{c|c}
	\bm{W} \diag(e^{i \bm{\delta}} \tanh \bm{r}) \bm{W}^T &  	-\bm{W} \diag(\sech  \bm{r}) \bm{V}\\
	\hline
	-\bm{V}^T \diag(\sech  \bm{r}) \bm{W^T} & - \bm{V}^T \diag(e^{-i \bm{\delta}}\tanh \bm{r}) \bm{V}\label{Eq:Sigma}
\end{array}
\right],
\end{align}
where $\bm\gamma$ is the vector of displacement parameters, $\bm \delta$ and $\bm r$ are the polar coordinates of the complex vector $\bm \zeta$ of squeezing parameters, and $\bm W$ and $\bm V$ are unitary covariance matrices describing the two interferometers.

\section{Generating the transformed state directly}
\subsection{Single-mode}
The matrix of a single-mode Gaussian transformation has rank 2. According to Eq.~\eqref{recrel}, the first row is given by:
\begin{align}
    \label{eq:G0}
    \mathcal{G}_{0,n} = \frac{\mu_2}{\sqrt{n}}\mathcal{G}_{0,n-1} - \Sigma_{22}\sqrt{\frac{n-1}{n}}\mathcal{G}_{0,n-2},
\end{align}
where the first element is $\mathcal{G}_{0,0} = C$.

We indicate the first row in Dirac notation as $\langle\mathcal{G}_{0}|$.
By using \eqref{recrel} again, each successive row of the Gaussian transformation can be computed recursively:
\begin{align}
    \label{eq:Gm}
    \langle\mathcal{G}_{m}| = \frac{\mu_1}{\sqrt{m}}\langle \mathcal{G}_{m-1}| - \Sigma_{11}\sqrt{\frac{m-1}{m}}\langle \mathcal{G}_{m-2}| - \frac{\Sigma_{12}}{\sqrt{m}}\langle \mathcal{G}_{m-1}|a,
\end{align}
where $a$ is the annihilation operator.

Given an input state $|\psi\rangle$, the $m$-th amplitude of the output state is given by $\langle \mathcal{G}_m|\psi\rangle$.
Therefore, using the recurrence relation for the rows $\langle\mathcal{G}_m|$ in Eq.~\eqref{eq:Gm}, we can write:
\begin{align}
    \underbrace{\langle \mathcal{G}_m|\psi\rangle}_{R_m^{(0)}} =& \frac{\mu_1}{\sqrt{m}}\underbrace{\langle \mathcal{G}_{m-1}|\psi\rangle}_{R_{m-1}^{(0)}} - \Sigma_{11}\sqrt{\frac{m-1}{m}}\underbrace{\langle \mathcal{G}_{m-2}|\psi\rangle}_{R_{m-2}^{(0)}}\nonumber \\
    &- \frac{\Sigma_{12}}{\sqrt{m}}\underbrace{\langle \mathcal{G}_{m-1}|a|\psi\rangle}_{R_{m-1}^{(1)}},
\end{align}
where we have defined $R_m^{(k)} = \langle \mathcal{G}_m|a^k|\psi\rangle$.

This relation must hold for any $k \geq 0$:
\begin{align}
    \label{eq:fullRm}
    R_m^{(k)} = \frac{\mu_1}{\sqrt{m}}R_{m-1}^{(k)} - \Sigma_{11}\sqrt{\frac{m-1}{m}}R_{m-2}^{(k)} - \frac{\Sigma_{12}}{\sqrt{m}}R_{m-1}^{(k+1)},
\end{align}
which means that we can write all the final amplitudes $R_m^{(0)}$ as linear combinations of elements in $R_0^{(k)}$.

\begin{figure}[H] 
\begin{center}
\includegraphics[scale = 0.5]{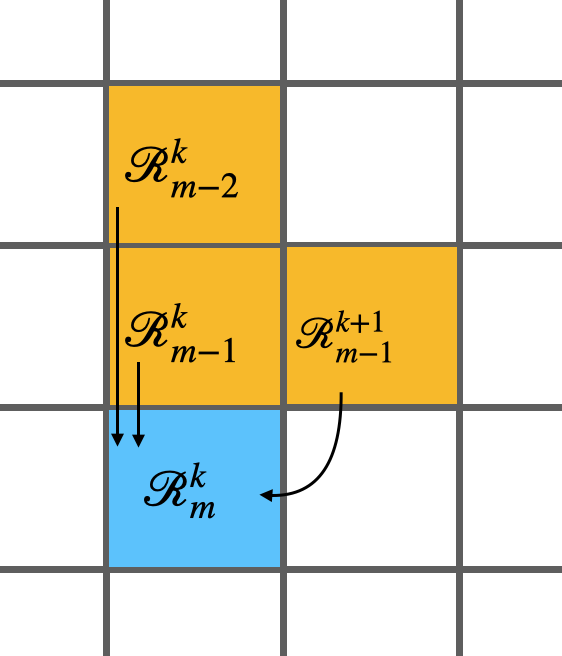}
\end{center}
\caption{Any element in the $R$ matrix (e.g.~the blue one at the bottom) can be calculated as a linear combination of the three neighbouring matrix elements shown in orange.}
\label{fig:singlemode_neighbour_illu}
\end{figure}

Imagine now $R_{m}^{(k)}$ as a matrix whose rows are indexed by $m$ and whose columns are indexed by $k$. From Eq.~\eqref{eq:fullRm} we know that each element in $R$ only depends on its three neighbouring elements (see Fig.~\ref{fig:singlemode_neighbour_illu}).

As we are interested in the first column of $R$, we start by filling the first row $R_0^{(k)}$ and then apply the rule iteratively, stopping at one fewer element per row, effectively filling half of the matrix.  This discount in computation carries over to each mode, i.e., for $M$ modes we need to compute only a fraction $2^{-M}$ of elements. As we need to combine $2M+1$ numbers to compute each new element of $R$, the time complexity of our algorithm is $O(MN^{2M}/2^M)$, which is exponentially faster in the number of modes with respect to computing the whole tensor $\mathcal G$, which has complexity $O(MN^{2M})$. 

In our Python code, we proceed as follows:
\begin{enumerate}
    \item We build a vector of square roots of integers up to $N$ to avoid having to recompute them often.
    \item We build the first row of the transformation matrix, $\langle \mathcal{G}_0|$.
    \item We build the first row $R_0^{(k)}$ by computing $\langle\mathcal{G}_{0}|a^k|\psi\rangle$ for $k\leq N$. To do so we take inner products only between the relevant part of the vector $\langle \mathcal{G}_0|$ and we update the state as $a^k|\psi\rangle$ for the next value of $k$.
    \item Starting from $m=1$, we compute each row $R_m^{(k)}$ using Eq.~\eqref{eq:fullRm}, but we only need $N-m$ elements.
    \item After we have built the last row $R_N^{(0)}$ (consisting of a single element), we read out the final result from the first column $R_m^{(0)}$.
\end{enumerate}

The code can then be sped-up via the Numba \cite{lam2015numba} decorator \verb|numba.njit|, and it will be compiled at the first call:

\begin{minted}[bgcolor=LightGray]{python}
@njit
def new_state(C, mu, Sigma, state):
  # 1. Setup
  N = len(state)
  dtype = state.dtype
  sqrt = np.sqrt(np.arange(N, dtype=dtype))

  # 2. First row of the transformation matrix
  G = np.zeros(N, dtype=dtype)
  G[0] = C
  for n in range(1, N):
    G[n] = (mu[1]/sqrt[n]*G[n-1] -
    Sigma[1,1]*sqrt[n-1]/sqrt[n]*G[n-2])

  # 3. First row of the R matrix
  R = np.zeros((N, N), dtype=dtype)
  for k in range(N):
    R[0,k] = np.dot(G[:N-k], state)
    state = state[1:]*sqrt[1:N-k]

  # 4. Other rows of the R matrix
  for m in range(1, N):
    for k in range(N-m): 
      R[m, k] = (mu[0]*R[m-1, k]/sqrt[m] 
      -Sigma[0,0]*R[m-2, k]*sqrt[m-1]/sqrt[m]
      -Sigma[0,1]*R[m-1, k+1]/sqrt[m])
  
  # 5. Readout
  return R[:, 0]
\end{minted}
\subsubsection{Special case: large squeezing}
If the squeezing parameter is large we don't need to generate the whole $R$ matrix to find the transformed state. To see that this is the case, notice that in \eqref{Eq:Sigma} we can approximate $\sech(r)\sim 0$ and $\tanh(r)\sim 1$ as $\mathbf{r}$ grows. We do maintain $\sqrt{\sech(r)}$ in $C$, as it is quadratically larger than $\sech{r}$.

For a single-mode Gaussian transformation, we have:
\begin{align}
    C &\sim \sqrt{\sech r} \exp\left(-\frac12|\gamma|^2-\frac12{\gamma^*}^2e^{i(\delta+2\phi)}\right),\\
    \bm \mu^T &\sim [\gamma^*e^{i(\delta+2\phi)}+\gamma,0],\\
    \bm{\Sigma} &\sim
    \left[	\begin{array}{c|c}
	 e^{i(\delta+2\phi)}   &  	0\\
	\hline
	0 & -e^{-i \delta}
\end{array}
\right]. 
\end{align}

We can hence rewrite the recurrence relations of Eq.~\eqref{eq:G0} and \eqref{eq:fullRm} for a single-mode Gaussian transformation with large squeezing:
\begin{align}
\mathcal{G}_{0,n} &\sim e^{-i\delta}\sqrt{\frac{n-1}{n}}\mathcal{G}_{0,n-2}\\
    \label{eq:fullRm_largesqueezing}
    R_m^{(0)} &\sim \frac{\gamma^*e^{i(\delta+2\phi)}+\gamma}{\sqrt{m}}R_{m-1}^{(0)} - e^{i(\delta+2\phi)}\sqrt{\frac{m-1}{m}}R_{m-2}^{(0)} 
\end{align}
Notice that in the second relation, $R^{(0)}_m$ does not depend on $R^{(k)}_m$ for $k>0$. We then use the first relation to compute $R_0^{0}$ and the second to compute the output state directly. We remark that under these conditions one can (differentiably) approximate the transformed state up to a cutoff in the order of thousands in milliseconds.  

\begin{figure}[!h]
    \centering
    \includegraphics[scale=0.6]{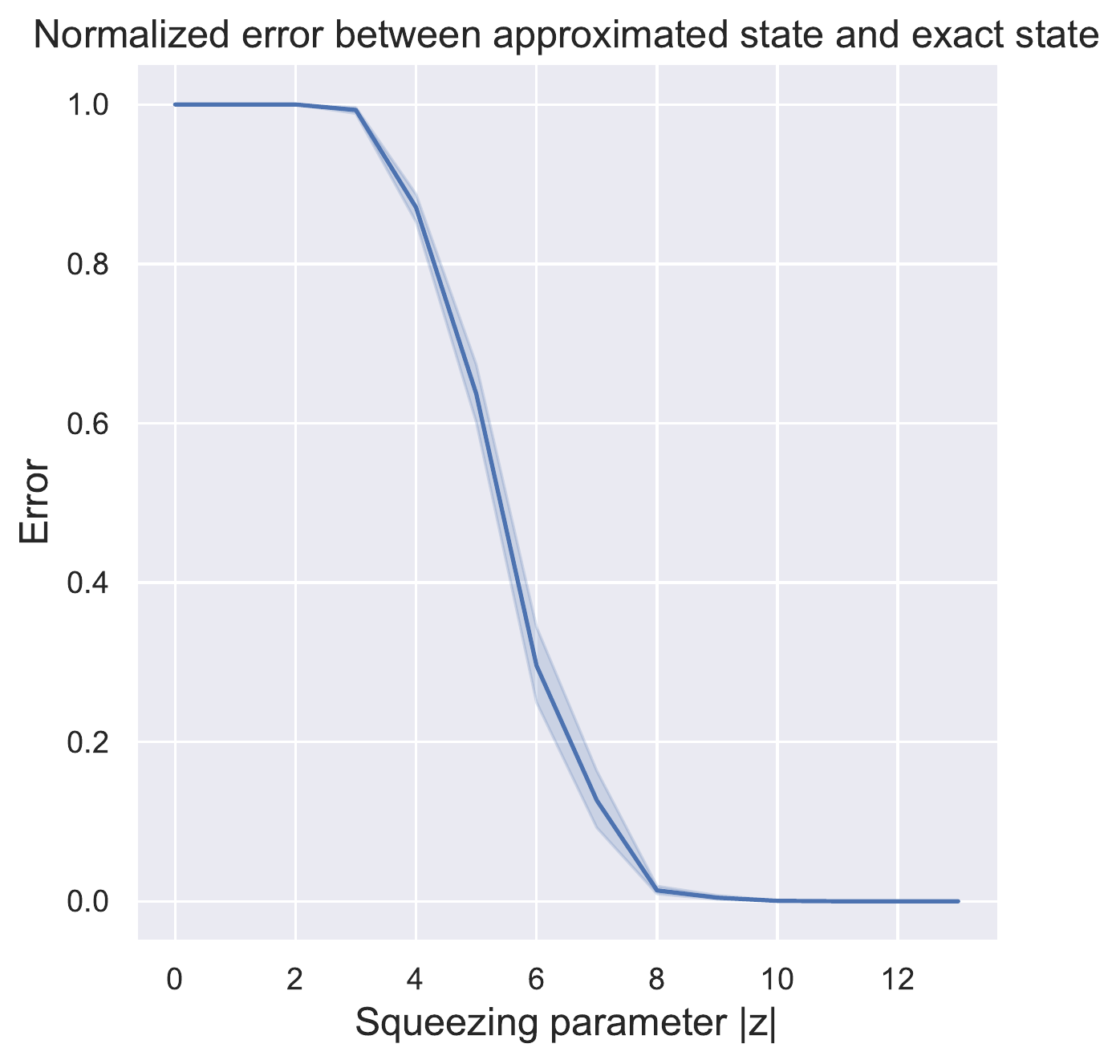}
    \caption{The error in the the normalized overlap between the approximated state and the exact state (here a set of 10 random states) goes to zero as the magnitude of the squeezing parameter $r = |z|$ increases. Note that the overlap only includes photon number states up to $n=50$.}
    \label{fig:error}
\end{figure}

To assess the quality of the approximation within the cutoff, in Fig.~\ref{fig:error} we show the deviation from the normalized overlap with respect to the exact state as a function of the parameter $r$ of the squeezing for a few random states.
Note that we need the renormalization because we compute the exact output state only up a fixed cutoff dimension, but large squeezing populates also Fock states with very large photon numbers.

\subsection{Two-mode Gaussian transformed state}
The transformation operator of a two-mode Gaussian transformation has rank 4, so we will build its elements by using our recursive equations four times.

The first two indices are filled by:
\begin{align}
    \mathcal{G}_{0,0,0,q}  &= \frac{1}{\sqrt{q}} \left( 
    \mathcal{G}_{0,0,0,q-1}\mu_{4} 
    - \sqrt{q-1}\mathcal{G}_{0,0,0,q-2}\Sigma_{44}
    \right),
\label{Eq:2mode_1}\\
    \mathcal{G}_{0,0,p,q}  &= \frac{1}{\sqrt{p}} ( 
    \mathcal{G}_{0,0,p-1,q}\mu_{3} 
    - \sqrt{p-1}\mathcal{G}_{0,0,p-2,q}\Sigma_{33} \nonumber \\
    &- \sqrt{q}\mathcal{G}_{0,0,p-1,q-1}\Sigma_{34}).
\label{Eq:2mode_2}
\end{align}

Using Dirac notation $\langle\mathcal{G}_{m,n}|$ we can express the following identities:
\begin{align}
    \mathcal{G}_{m,n,p-1,q}\sqrt{p} &= \langle\mathcal{G}_{mn}|a,\\
    \mathcal{G}_{m,n,p,q-1}\sqrt{q} &= \langle\mathcal{G}_{mn}|b,
\end{align}
where $a$ and $b$ are annihilation operators in the first and second mode, respectively.

We then apply the recurrence relation again to fill the third and fourth indices:
\begin{align}
    \langle\mathcal{G}_{0,n}|  &= \frac{1}{\sqrt{n}} ( 
    \mu_{2} \langle\mathcal{G}_{0,n-1}|
    - \Sigma_{22}\sqrt{n-1}\langle\mathcal{G}_{0,n-2}|\nonumber \\
    &\ - \Sigma_{23}\langle\mathcal{G}_{0,n-1}|a
    - \Sigma_{24}\langle\mathcal{G}_{0,n-1}|b),
\label{Eq:G0n}\\
    \langle\mathcal{G}_{m,n}| &= \frac{1}{\sqrt{m}} ( 
    \mu_{1} \langle\mathcal{G}_{m-1,n}|
    - \Sigma_{11}\sqrt{m-1}\langle\mathcal{G}_{m-2,n}|\nonumber \\
    &\ - \Sigma_{12}\sqrt{n}\langle\mathcal{G}_{m-1,n-1}|
    - \Sigma_{13}\langle\mathcal{G}_{m-1,n}|a \nonumber\\
    &\ - \Sigma_{14}\langle\mathcal{G}_{m-1,n}|b).
\label{Eq:Gmn}
\end{align} 

We now assume a pure two-mode input state $|\bm{\psi}\rangle$ and we compute the $(m,n)$-th amplitude of the output state: $\langle \mathcal{G}_{mn}|\bm{\psi}\rangle$. Using the recurrence relation in Eq.~\eqref{Eq:G0n} and Eq.~\eqref{Eq:Gmn}, and defining $R_{m,n}^{(j,k)} = \langle\mathcal{G}_{m,n}|a^j b^k |\bm{\psi}\rangle$, we get:
\begin{align}
    \underbrace{\langle\mathcal{G}_{0,n}|\bm{\psi}\rangle}_{R_{0,n}^{(0,0)}}
    &= \frac{1}{\sqrt{n}} 
    (
    \mu_{2} \underbrace{\langle\mathcal{G}_{0,n-1}|\bm{\psi}\rangle}_{R_{0,n-1}^{(0,0)}} - \Sigma_{22}\sqrt{n-1}\underbrace{\langle\mathcal{G}_{0,n-2}|\bm{\psi}\rangle}_{R_{0,n-1}^{(0,0)}} \nonumber \\
    &- \Sigma_{23}\underbrace{\langle\mathcal{G}_{0,n-1}|a|\bm{\psi}\rangle}_{R_{0,n-1}^{1,0}}
    - \Sigma_{24}\underbrace{\langle\mathcal{G}_{0,n-1}|b|\bm{\psi}\rangle}_{R_{0,n-1}^{0,1}}
    ),
\label{eq:R0n1}\\
    \underbrace{\langle\mathcal{G}_{m,n}|\bm{\psi}\rangle}_{R_{m,n}^{0,0}} &= \frac{1}{\sqrt{m}} 
    (
    \mu_{1} \underbrace{\langle\mathcal{G}_{m-1,n}|\bm{\psi}\rangle}_{R_{m-1,n}^{0,0}}
    - \Sigma_{11}\sqrt{m-1}\underbrace{\langle\mathcal{G}_{m-2,n}|\bm{\psi}\rangle}_{R_{m-2,n}^{0,0}}\nonumber \\
    &- \Sigma_{12}\sqrt{n}\underbrace{\langle\mathcal{G}_{m-1,n-1}|\bm{\psi}\rangle}_{R_{m-1,n-1}^{0,0}}
    - \Sigma_{13}\underbrace{\langle\mathcal{G}_{m-1,n}|a|\bm{\psi}\rangle}_{R_{m-1,n}^{1,0}}\nonumber \\
    &- \Sigma_{14}\underbrace{\langle\mathcal{G}_{m-1,n}|b|\bm{\psi}\rangle}_{R_{m-1,n}^{0,1}}
    ).
\label{eq:Rmn2}
\end{align} 

The relation holds for any $j$ and $k$, so we finally obtain:
\begin{align}
    R_{0,n}^{(j,k)} 
    &= \frac{1}{\sqrt{n}} ( 
    \mu_{2} R_{0,n-1}^{(j,k)}
    - \Sigma_{22}\sqrt{n-1}R_{0,n-2}^{(j,k)}\nonumber\\
    &- \Sigma_{23}R_{0,n-1}^{(j+1,k)}
    - \Sigma_{24}R_{0,n-1}^{(j,k+1)}),\label{eq:R0n}\\
    R_{m,n}^{(j,k)} 
    &= \frac{1}{\sqrt{m}} ( 
    \mu_{1} R_{m-1,n}^{(j,k)}
    - \Sigma_{11}\sqrt{m-1}R_{m-2,n}^{(j,k)}\nonumber \\
    &- \Sigma_{12}\sqrt{n}R_{m-1,n-1}^{(j,k)}
    - \Sigma_{13}R_{m-1,n}^{(j+1,k)}
    - \Sigma_{14}R_{m-1,n}^{(j,k+1)}).\label{eq:Rmn}
\end{align} 

Similarly to the single-mode case, because of the dependencies among neighboring elements, we compute the output state without needing to generate the whole rank-4 tensor. 

In our python code for the two-mode case we proceed as follows :
\begin{enumerate}
    \item We build a vector of square roots of integers up to $N$ to avoid having to recompute them often.
    \item We compute the matrix $\langle \mathcal{G}_{00}|$ using Eq.~\eqref{Eq:2mode_1} and \eqref{Eq:2mode_2}.
    \item We compute $R_{00}^{(j,k)}$ by multiplying $\langle\mathcal{G}_{00}|a^j b^k$ and $|\bm\psi\rangle$. This step is analogous to the single-mode case except we loop over two indices instead of one.
    \item We compute $R_{0n}^{(j,k)}$ using Eq.~\eqref{eq:R0n} and $R_{mn}^{(j,k)}$ using Eq.~\eqref{eq:Rmn}. We do this under the conditions: $m< N,\ n < N,\ j< N-m,\ k< N-m-j$, which allow us to compute only one quarter of the possible $N^4$ elements of $R_{mn}^{(j,k)}$.
    \item We read out the final result from $R_{mn}^{(0,0)}$.
\end{enumerate}

One can extend this method to any number of modes by following a similar procedure with a larger number of indices. In the $M$-mode case, the conditions will allow one to obtain the final state by computing only a fraction of $2^{-M}$ elements of the rank $2M$ tensor $R$.

\section{Gradients of the transformed state}
The recurrence relations that we have presented are all linear, and therefore they can be easily differentiated. This gives us a direct way of computing the gradient of the transformed state with respect to the parameters of the transformation.

In our Gaussian transformations we use a mix of real and complex parameters. We will therefore follow Wirtinger calculus to compute the gradients with respect to complex parameters: Wirtinger derivatives are computed by considering a complex variable $\xi$ and its complex conjugate $\xi^*$ as \emph{independent} variables. This allows us to compute derivatives of non-holomorphic functions, which can be thought of as functions that explicitly depend on $\xi^*$ such as the absolute value $|\xi| = \sqrt{\xi\xi^*}$. If a function is holomorphic, its derivative with respect to $\xi^*$ is zero and in this case the derivative with respect to $\xi$ is sufficient. This is a consequence of the fact that holomorphic functions satisfy the  Cauchy-Riemann equations, establishing a dependence between the derivatives of the real and imaginary parts of the function and thereby removing a degree of freedom.

Assuming we have defined a real loss function $L:\mathbb{C}^n \rightarrow \mathbb{R}$ that maps the output state of the Gaussian transformation to a real value, we can choose to use gradient descent to optimize it. Note that for a complex parameter $\xi$, the correct gradient to utilize is the one with respect to the \emph{conjugate} of the parameter we want to optimize, i.e., the gradient update rule reads:
\begin{align}
    \label{eq:complexupdate}
    \xi \leftarrow \xi - \epsilon\frac{\partial L}{\partial \xi^*}
\end{align}
for a learning rate $\epsilon$. This gradient update rule falls back to the usual one in case $\xi$ is a real variable.

We consider quantum transformations that depend on a complex parameter $\xi$ as functions $f:\mathbb{C}^n\times\mathbb{C}\rightarrow \mathbb{C}^n$, such that $\bm \psi^{(out)} = f(\bm \psi^{(in)}, \xi)$.
Therefore, we may want to backpropagate the upstream gradient $\partial L/\partial\bm\psi^{(out)}$ to the gradient with respect to the parameter $\partial L/\partial\xi^*$ or to the gradient with respect to the input $\partial L/\partial\bm\psi^{(in)}$.
As we need to treat any complex variable and its conjugate as independent, we need to consider both when applying the chain rule. 

For our code, we use the automatic differentiation framework of TensorFlow, which computes the gradient $\frac{\partial L}{\partial \psi_{\bm k}^*}$ for us, where $\psi_{\bm k}^*$ is the conjugate of the input to the loss function (note that TensorFlow computes the appropriate gradient as per Eq.~\eqref{eq:complexupdate}). From this gradient we obtain $\frac{\partial L}{\partial \psi_{\bm k}} = \left(\frac{\partial L}{\partial \psi_{\bm k}^*}\right)^*$ by conjugation.

Then, if we want to backpropagate the upstream gradient to the input of the previous transformation, we have to provide the conjugate of the transformation itself: $\frac{\partial {\psi^{(out)}}^*_{\bm k}}{\partial {\psi^{(in)}}^*_{\bm k'}} = \mathcal{G}_{{\bm k},{\bm k}'}^*$:
\begin{align}
    \frac{\partial L}{\partial {\psi^{(in)}}_{\bm k'}^*} &= \sum_{\bm k} \frac{\partial L}{\partial {\psi^{(out)}_{\bm k}}^*}\frac{\partial {\psi^{(out)}}_{\bm k}^*}{\partial {\psi^{(in)}}_{\bm k'}^*} + \frac{\partial L}{\partial {\psi^{(out)}_{\bm k}}}\frac{\partial {\psi^{(out)}}_{\bm k}}{\partial {\psi^{(in)}}_{\bm k'}^*}\nonumber\\
    &=\sum_{\bm k} \frac{\partial L}{\partial {\psi^{(out)}_{\bm k}}^*}\mathcal{G}_{{\bm k},{\bm k}'}^*,
\end{align}
the second term being always zero.

If instead we want to backpropagate the upstream gradient to the parameter of the transformation, we have to provide $\frac{\partial \psi_k}{\partial \xi^*}$ and $\frac{\partial \psi_k^*}{\partial \xi^*} = \left(\frac{\partial \psi_k}{\partial \xi}\right)^*$:
\begin{align}
    \frac{\partial L}{\partial \xi^*} = \sum_{\bm k} \frac{\partial L}{\partial \psi_{\bm  k}^*}\frac{\partial \psi_{\bm k}^*}{\partial \xi^*} + \frac{\partial L}{\partial \psi_{\bm k}}\frac{\partial \psi_{\bm k}}{\partial \xi^*}.
\end{align}

To compute the desired gradients (here for a single-mode), we differentiate Eq.~\eqref{eq:Gm} and Eq.~\eqref{eq:fullRm} with respect to a parameter $\xi$ (or its conjugate $\xi^*$):
\begin{align}
    \partial_\xi \mathcal{G}_{0,n} = & \frac{1}{\sqrt{n}}\left[\mu_2(\partial_\xi \mathcal{G}_{0,n-1}) +(\partial_\xi \mu_2) \mathcal{G}_{0,n-1} \right] \nonumber \\
    &- \sqrt{\frac{n-1}{n}}\left[
    (\partial_\xi \Sigma_{22})\mathcal{G}_{0,n-2} + \Sigma_{22}(\partial_\xi \mathcal{G}_{0,n-2})
    \right],\\
    \partial_\xi R_m^{(k)} = & \frac{1}{\sqrt{m}}\left[(\partial_\xi\mu_1)R_{m-1}^{(k)} + \mu_1(\partial_\xi R_{m-1}^{(k)})\right]\nonumber\\
    &- \sqrt{\frac{m-1}{m}}\left[(\partial_\xi\Sigma_{11})R_{m-2}^{(k)} + \Sigma_{11}(\partial_\xi R_{m-2}^{(k)})\right]\nonumber \\
    &-\frac{1}{\sqrt{m}}\left[(\partial_\xi\Sigma_{12})R_{m-1}^{(k+1)} + \Sigma_{12}(\partial_\xi R_{m-1}^{(k+1)})\right].
\end{align}

And we obtain the gradient from the first column:
\begin{align}
    \frac{\partial \psi_m}{\partial \xi} = \partial_\xi R_m^{(0)}.
\end{align}
These recurrence relations can be computed once the $R$ matrix is known. Therefore, after generating the matrix $R$, we keep it in memory for the computation of the gradient. To speed things up even further, we pre-compute and compile the derivative functions of $C$, $\bm \mu$ and $\bm \Sigma$ with respect to the parameters.

From a practical point of view, the backpropagated gradients have the same shape of the state (i.e., $N^M$ for an $N$-dimensional Fock space on $M$-modes) rather than the shape of the transformation (i.e., $N^{2M}$), which is a quadratic advantage over previous approaches. We still have to compute the transformation itself in order to backpropagate the gradients, but instead of the gradient of the transformation we can just compute the gradient of $R$, which is only a fraction $2^{-M}$ in size.

\section{Numerical experiments}
In this section we present benchmarks of various experiments based on our new method.

\begin{figure*}
    \centering
    \includegraphics[scale=0.6]{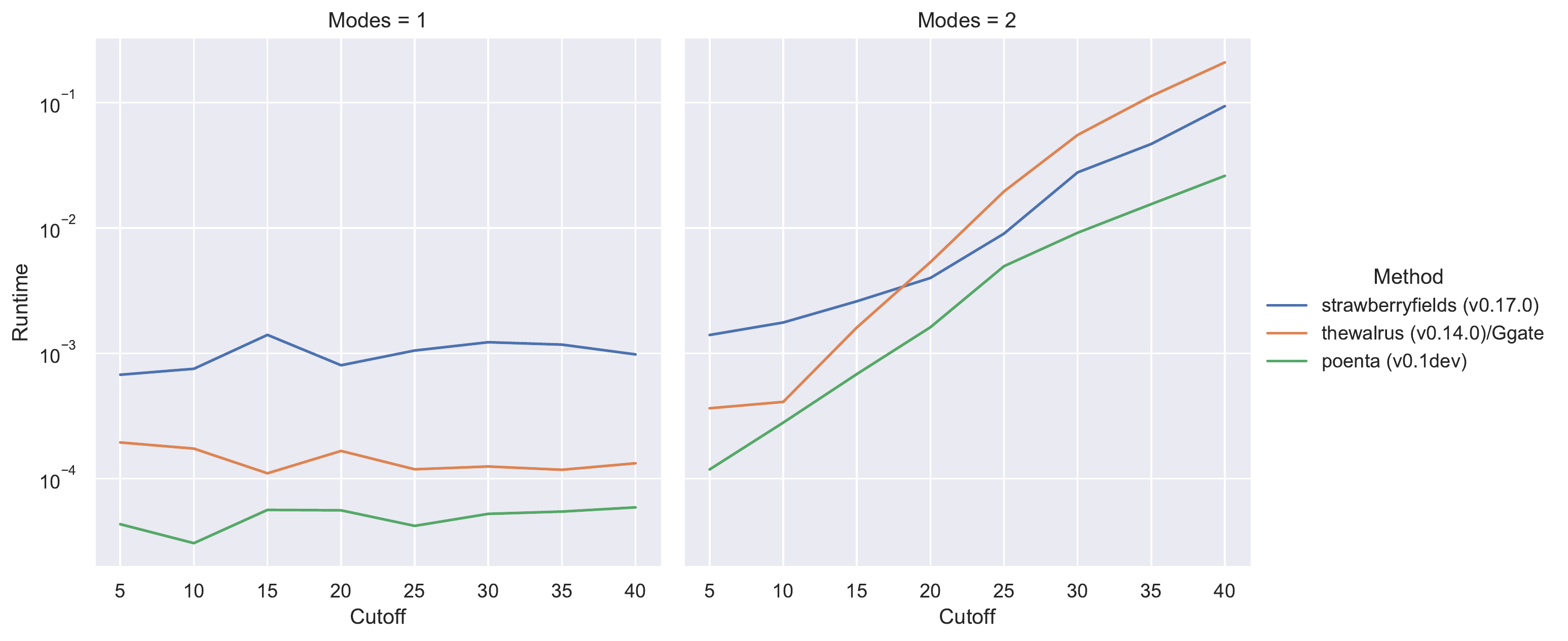}
    \caption{Our implementation (lowest line, green) is up to an order of magnitude faster at generating the output of a Gaussian transformation than the other methods. For the two-mode case, Strawberry Fields has the advantage of exploiting the photon-number conservation of the beamsplitter to spare one nested loop. In the other two implementations we build the tensors all at once, and therefore we don't have access to the same savings. Despite this, our current implementation is still faster and even more so when the backward pass is computed (see Table I), as the backpropagated gradients have the shape of the state rather than the shape of the transformation.}
    \label{fig:comp3}
\end{figure*}

First, we benchmark the runtime to compute the transformed state (i.e., the forward pass) against the current state of the art. We obtain the transformed state by three methods: constructing the Gaussian transformation matrix by using a sequence of gates in StrawberryFields 0.17 \cite{killoran2019strawberry} and taking the matrix-vector product, by getting the Gaussian transformation matrix from our previous work in the Ggate branch of The Walrus \cite{gupt2019walrus} and taking the matrix-vector product, and with the method in this work to get the transformed state directly. The results are summarized in Fig.~\ref{fig:comp3}.

Second, we benchmark a circuit optimization task for various one- and two-mode states  (i.e., we benchmark both the forward and the backward pass). 
Our architecture is made of a sequence of $L$ layers, each formed by a Gaussian gate followed by a single-mode Kerr gate in each mode:
\begin{align}
    U(\bm \beta) = \prod_{\ell=1}^L \mathcal{G^{\ell}(\bm{\alpha}_{\ell})}K(\bm\kappa_\ell),
\end{align}
where $\bm \alpha_{\ell}$ denotes the parameters of the $\ell$-th Gaussian layer and $\bm \beta$ denotes the parameters of the whole circuit.
Each single-mode Kerr transformation is given by
\begin{align}
    K(\kappa) = e^{i\kappa (a^\dagger a)^2},
\end{align}
where $\kappa$ is a positive parameter.

Suppose we have a set of $S$ required input and target pairs $\{(\Psi^\mathrm{in}_s,\Psi^\mathrm{target}_s)\}$,
our goal is to find a circuit that can satisfy the requirements. The optimization is performed by minimizing the loss function:
\begin{align}
    L(\bm\beta) = 1 - \frac{1}{S}\sum_{s=1}^{S}|\langle \Psi^\mathrm{target}_s|U(\bm \beta)|\Psi^\mathrm{in}_s\rangle|^2.
\end{align}

Note that if the number of required input-output pairs is large, the optimizer might content itself with satisfying only a few of them. In that case, we can include in the loss function a term that favours a uniform probability over all the required input-output pairs, for example the KL divergence with respect to the uniform distribution: $-\sum_s\log|\langle \Psi^\mathrm{target}_s|U(\bm \beta)|\Psi^\mathrm{in}_s\rangle|^2$, which diverges rapidly as soon as either of the probabilities tries to go to zero.

We ran benchmarks on generating a single-photon ($|\Psi^\mathrm{in}\rangle = |0\rangle, |\Psi^\mathrm{target}\rangle = |1\rangle$), NOON state with $N=5$ ($|\Psi^\mathrm{in}\rangle = |0,0\rangle, |\Psi^\mathrm{target}\rangle = (|5,0\rangle+|0,5\rangle)/\sqrt{2}$) and Hex GKP state as in \cite{arrazola2019machine} with $d=2$, $\mu=1$, $\delta=0.3$. The results are shown in the table \ref{tab:Stategeneration}. One can compare them with our previous implementation in table I of \cite{miatto2020fast} and with the original implementation in table I of \cite{arrazola2019machine}, while taking into account differences in computational hardware.

\begin{table}[h]
    \centering
    \begin{tabular}{c|c|c|c}
    \hline
    \hline
     Hyperparams    &  Single photon & NOON state & Hex GKP \\
    \hline
     Modes M    &  1  & 2 & 1\\
     Cutoff N & 100  & 10  & 50\\
     Layers L & 8  &  20 & 25 \\
     Steps & 1500   &  3000  & 10000\\
    \hline
    \hline
     Results    &  & &\\
    \hline
    Fidelity &  99.992\%  &  99.905\% & 99.537\%\\
    Runtime(s) & 24s & 341s & 410s\\
    \hline
    \end{tabular}
    \caption{Runtime for three circuit optimization tasks (single-core on an Apple M1 chip). We achieve about twice the performance as compared to our previous implementation, which was more than one order of magnitude faster than previous methods.}
    \label{tab:Stategeneration}
\end{table}

\section{Conclusions}

In this work we introduced a new algorithm to compute how a quantum state evolves under a Gaussian transformation. In comparison to previous approaches, our algorithm has an advantage in runtime and memory that is exponential in the number of optical modes. Moreover it is differentiable, which means we can  calculate gradients of the resulting state with respect to the transformation parameters and use them to train or optimize quantum optical circuits.
We made available our algorithm through our library \emph{poenta} \cite{poenta}.

The main outstanding challenge of our work is that we assume the input state is pure. If the input state is mixed, we would have to run the algorithm on each significant eigenstate of the density matrix, thus getting a performance hit proportional to the rank of the state. However, in preliminary numerical benchmarks we have noticed no impact, probably thanks to automatic code vectorization. We will investigate this avenue in future work.

\bibliography{fast_evolution}

\end{document}